\documentclass[
superscriptaddress,
amsmath,amssymb,
aps, %pra,
prl,
twocolumn, 
floatfix, 
longbibliography
]{revtex4-2}
\usepackage{graphicx}
\usepackage{siunitx}
\usepackage{amsmath}
\usepackage{hyperref}
\usepackage[nameinlink,capitalize]{cleveref}
\usepackage{breakurl}   % if using dvips
\usepackage{xurl} 
\usepackage{soul}
\usepackage{longtable}
\usepackage{xr}
\usepackage{cancel}
\usepackage[normalem]{ulem}
\usepackage{pdfpages}
\makeatletter
\AtBeginDocument{\let\LS@rot\@undefined}
\makeatother
\usepackage{pgffor}
\crefname{figure}{Fig.}{Figs.}
\Crefname{figure}{Figure}{Figures}

\crefname{table}{Table}{Tables}
\Crefname{table}{Table}{Tables}

\crefname{equation}{Eq.}{Eqs.}
\Crefname{equation}{Equation}{Equations}

\crefname{section}{Sec.}{Secs.}
\Crefname{section}{Section}{Sections}

\crefname{SIfigure}{Fig.~S}{Figs.~S}
\Crefname{SIfigure}{Figure~S}{Figures~S}

\crefname{SItable}{Table~S}{Tables~S}
\Crefname{SItable}{Table~S}{Tables~S}

\crefname{SIequation}{Eq.~S}{Eqs.~S}
\Crefname{SIequation}{Equation~S}{Equations~S}

\crefname{SIsection}{Sec.~SI}{Secs.~SI}
\Crefname{SIsection}{Section~SI}{Sections~SI}

\begin{document}

\title{Magnon Kerr effect in a magnetic thin film strongly coupled to a microwave resonator}

\author{Davit Petrosyan}
\email[]{davit.petrosyan@mat.ethz.ch}
%\thanks{}
\affiliation{Department of Materials, ETH Zurich, CH-8093 Zurich, Switzerland}
\author{Hiroki Matsumoto}
%\thanks{}
\affiliation{Department of Materials, ETH Zurich, CH-8093 Zurich, Switzerland}
\affiliation{Institute for Chemical Research, Kyoto University, 6110011 Uji, Japan}
\author{Hanchen Wang}
%\thanks{}
\affiliation{Department of Materials, ETH Zurich, CH-8093 Zurich, Switzerland}
\author{Jamal Ben~Youssef}
%\thanks{}
\affiliation{LabSTICC, CNRS, Universit{\'{e}} de Bretagne Occidentale, 29238 Brest, France}
\author{Richard Schlitz}
%\thanks{}
\affiliation{Department of Physics, University of Konstanz, 78457 Konstanz, Germany}

\author{William Legrand}
\email[]{william.legrand@neel.cnrs.fr}
%\thanks{}
\affiliation{Department of Materials, ETH Zurich, CH-8093 Zurich, Switzerland}
\affiliation{Université Grenoble Alpes, CNRS, Institut Néel, 38042 Grenoble, France}

\author{Pietro Gambardella}
\email[]{pietro.gambardella@mat.ethz.ch}
%\thanks{}
\affiliation{Department of Materials, ETH Zurich, CH-8093 Zurich, Switzerland}

\begin{abstract}

Progress in cavity magnonics relies on tunable nonlinearities in magnon–photon interactions for controlling magnon dynamics and frequency conversion. Here, we demonstrate the magnon Kerr effect in anisotropic magnonic systems comprising iron garnet thin films coupled to a loop-gap microwave resonator. The strong shape anisotropy, combined with strain, enhances the magnon Kerr effect compared to a sphere of equivalent volume by up to three orders of magnitude, while the cavity enables sensitive probing of magnetization dynamics. We demonstrate continuous tunability of the magnitude and sign of the Kerr shift by controlling the static orientation of the magnetization, in excellent agreement with theory. Our findings demonstrate a scalable approach to enhancing Kerr anharmonicity in hybrid magnon--photon systems while preserving coherent magnon--photon interactions.

\end{abstract}

\maketitle

The field of cavity magnonics explores interactions between exchange-coupled electronic spins in magnetically ordered materials and microwave photons in electromagnetic resonators \cite{PhysRevLett.104.077202, Lachance-Quirion_2019, 10.1063/5.0020277,ZARERAMESHTI20221}. First demonstrated in experiments with single-crystal yttrium iron garnet (YIG) spheres \cite{PhysRevLett.113.083603, PhysRevLett.113.156401}, this concept has been extended to the ultra-strong coupling regime \cite{PhysRevLett.113.156401, PhysRevApplied.20.024039, yoshii2026onchipdicketypemagnonpolaritons}, planar cavities \cite{PhysRevLett.111.127003, PhysRevB.99.140414, PhysRevApplied.20.024039, Morris2017}, and integrated devices \cite{PhysRevLett.123.107701, PhysRevLett.123.107702, Guo2023}. By combining ferromagnetic resonance modes with cavity electrodynamics, strong magnon--photon coupling enables highly sensitive spectral probing of magnon dynamics. Furthermore, field confinement inside the cavity enhances the effective drive strength and prevents radiative decay, allowing for large magnon populations. This facilitates the exploration of nonlinear magnon regimes, where magnon–magnon interactions modify their collective response to microwave excitation \cite{SUHL1957209, PhysRev.100.1788, 809144, 10.1063/1.334997}.

The magnon Kerr effect (MKE) appears in all ferromagnets with finite magnetic anisotropy, manifesting as a self-induced frequency shift of the magnon modes. Such a shift has been observed in nonlinear ferromagnetic resonance measurements \cite{10.1063/1.327760, 10.1063/1.332537, 10.1063/1.334997} and cavity-coupled YIG spheres \cite{PhysRevB.94.224410}, where it gives rise to bistability of magnon--polaritons \cite{PhysRevLett.120.057202}. In contrast to nonlinear magnon processes driven by external parametric pumping \cite{PhysRevB.86.134420, BRACHER20171, 10.1063/5.0038946, PhysRevLett.99.037205}, the MKE does not require additional drive tones nor dispersion matching, such as three-magnon scattering processes \cite{SUHL1957209, Kurebayashi2011, PhysRevLett.103.157202, PhysRevLett.130.046703, PhysRevLett.85.2184, PhysRevB.67.104402, PhysRevB.79.144428, PhysRevLett.103.157202, jnpb-2mxx}. Moreover, Kerr nonlinearities can also arise between magnons and other bosonic quasi-particles \cite{PhysRevLett.129.123601, Shen2025, zw18-26nw} under sufficient populations of the respective modes. The MKE is hence a direct and desirable approach for inducing nonlinearities \cite{PhysRevA.111.013708, PhysRevA.110.043704, BHATT2025172275, https://doi.org/10.1002/qute.202400654, PhysRevApplied.12.034001} and entangle different degrees of freedom in hybrid magnonic systems \cite{PhysRevResearch.1.023021, 343d-t8pv, MA2025131897}.

In general, the MKE can be enhanced by reducing the volume of the magnonic system, or by increasing its magnetic anisotropy. This sets a limit to the MKE that can be achieved with YIG spheres, due to the lower bound on their volumes (diameter $>\SI{0.2}{mm}$) and their inherently small magnetocrystalline anisotropy. Micrometer- or nanometer-sized magnets strongly coupled to superconducting resonators \cite{PhysRevLett.123.107701, PhysRevLett.123.107702, Guo2023} provide an alternative solution, with magnetic volumes that are orders of magnitude smaller. However, the nonlinearity of the superconducting resonators made it difficult so far to disentangle the MKE from the cavity response \cite{PhysRevLett.123.107701}. These limitations motivate alternative approaches combining a millimeter-sized three-dimensional (3D) or a planar resonator with a thin magnetic film \cite{zw18-26nw,Jiang2026}, which can simultaneously enhance the MKE while maintaining a linear photonic probing.

\begin{figure}[t]
\includegraphics[width=\linewidth]{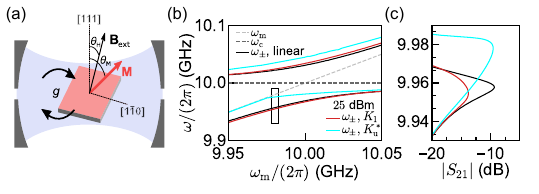}%
\caption{Magnon Kerr effect in a hybrid system based on a magnetic thin film and a microwave cavity. (a) The Kittel mode and cavity mode, with angular frequencies $\omega_\mathrm{m}$ and $\omega_\mathrm{c}$, couple at rate $g$. The magnetization and external field angles are denoted by $\theta_\mathrm{M}$ and $\theta_\mathrm{H}$, respectively. (b) Simulated hybridized magnon--polariton modes $\omega_{\pm}$ in the linear and nonlinear (\SI{25}{dBm}) regimes for magnetocrystalline ($K_1$) and uniaxial shape ($K_\mathrm{u}^*$) anisotropies. (c) Microwave transmission in the boxed region of b, showing the magnon Kerr shift of the lower polariton in the nonlinear regime. Simulation parameters are given in \cite{supp}.}
\label{fig:fig1_MKE}
\end{figure}

In this Letter, we report the MKE in magnetic insulator thin films strongly coupled to a 3D microwave cavity, schematically shown in \cref{fig:fig1_MKE}(a). The thin film geometry, combining reduced volume and strong shape anisotropy, significantly enhances the MKE compared to the smallest YIG spheres, leading to power-induced shifts of the magnon--polariton branches that can exceed the mode linewidth, as illustrated in \cref{fig:fig1_MKE}(b,c). To demonstrate this point and investigate the MKE originating from dipolar effects, we probe the magnon--polariton spectrum of thin film YIG as a function of microwave drive power. By varying the magnetization angle, we demonstrate control of the sign of the Kerr shift and continuous tunability of its strength. We further show that strain-induced magnetic anisotropy in thin films enhances the MKE, providing an additional degree of control. Our data are in excellent agreement with a recent analytical model of the MKE in thin films~\cite{petrosyan2026modelmagnonkerreffect}. These results establish the MKE of magnetic thin films as a superior approach to shape the nonlinear properties of cavity magnonic systems.

We investigate first a one-sided, \SI{200}{nm}-thick film of unstrained YIG grown on gadolinium gallium garnet (GGG) by liquid phase epitaxy, with the backside film polished away. The sample is cut to $\SI{2.5}{}\times\SI{3.5}{mm^2}$ and inserted into a loop-gap resonator~\cite{1mc9-k683} to form the magnon--photon system. The orientation of the uniform external field $\mu_0 H_\mathrm{ext}$ relative to the sample plane is continuously varied between in-plane (IP) and out-of-plane (OOP) by rotating the resonator in an electromagnet. The angles $\theta_\mathrm{H}$ and $\theta_\mathrm{M}$ refer to the applied external field and magnetization, respectively, from the normal of the film [\cref{fig:fig1_MKE}(a)]. Two ports of the resonator are connected to a vector network analyzer (VNA) to probe microwave transmission via the $S_{21}$ scattering matrix term. More details on the experimental setup and configuration are shown in the Supplemental Material (SM) \cite{supp}\nocite{Hansen1974a, https://doi.org/10.1002/adfm.202503644, 1mc9-k683, MAKINO1981957, 10.1063/1.4899186, KUBOTA201363, PhysRevB.89.134404, Soumah2018, https://doi.org/10.1002/admi.202300217, 8l57-yqsx, PhysRevB.95.214423}.
\begin{figure}[t]
\includegraphics[width=\linewidth]{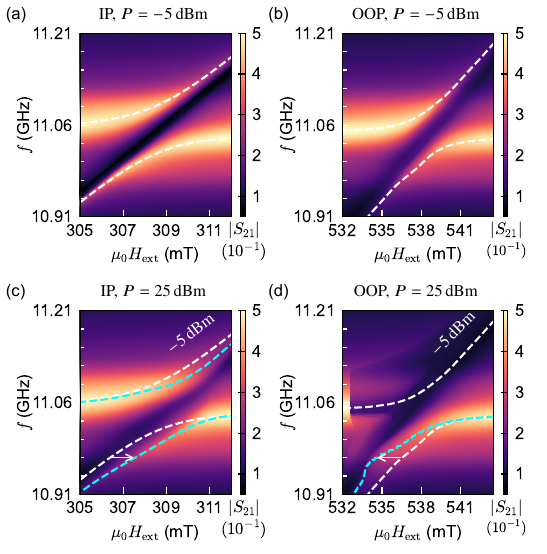}%
\caption{Field- and frequency-dependent microwave transmission of the coupled magnon–photon system in a YIG thin film in the IP configuration at (a) $P=\SI{-5}{dBm}$ and (c) $P = \SI{25}{dBm}$, and OOP configuration at (b) $P=\SI{-5}{dBm}$ and (d) $P = \SI{25}{dBm}$. Dashed lines follow the maxima in the transmission and mark the magnon–polariton branches at $P=\SI{-5}{dBm}$ (white) and $P=\SI{25}{dBm}$ (cyan). Arrows indicate the shift direction. The OOP upper magnon--polariton branch at $P = \SI{25}{dBm}$ does not appear due to foldover.}
\label{fig:fig2}
\end{figure}

We measure transmission in the IP configuration and linear excitation regime (microwave power $P=\SI{-5}{dBm}$), for which a colormap of $|S_{21}|$ as a function of magnetic field and microwave frequency is presented in \cref{fig:fig2}(a). The map features the avoided crossing between the cavity resonance and Kittel mode, corresponding to the hybridized magnon--photon polariton modes. The same measurement for OOP field at $P=\SI{-5}{dBm}$ is shown in \cref{fig:fig2}(b). We find the magnon--photon coupling constant $g/(2\pi)= \SI{35.9}{MHz}$, total cavity linewidth $\kappa_\mathrm{tot}/(2\pi)=\SI{64.0}{MHz}$, and magnon dissipation rates $\kappa_\mathrm{m}/(2\pi)=\SI{10.9}{MHz}$ and \SI{30.8}{MHz} for IP and OOP fields, respectively, from fits described in SM~\cite{supp}. In both configurations, we attain the limit of strong coupling regime $g>\kappa_\mathrm{tot}/2, \kappa_\mathrm{m}/2$ \cite{PhysRevLett.113.156401}. Notably, owing to our resonator design, the value for $g$ is comparable to that obtained in previous works with bulk YIG spheres, despite those spheres having a total number of spins several orders of magnitude larger \cite{PhysRevB.94.224410}. The cooperativity $4g^2/(\kappa_\mathrm{tot} \kappa_\mathrm{m})$, quantifying the coherence of the hybridization, is $7.3$ and $2.6$ for IP and OOP fields, respectively. The increase in $\kappa_\mathrm{m}$ for OOP field is mainly due to inhomogeneous broadening \cite{Soumah2018}.

The hybrid modes are then measured at $P=\SI{25}{dBm}$. For IP field [\cref{fig:fig2}(c)], the two polariton branches have shifted to higher fields (i.e., lower magnon frequencies), as visible by comparing transmission maxima (cyan) and reference (white) taken from \cref{fig:fig2}(a). For OOP field [\cref{fig:fig2}(d)], the lower branch has shifted in the opposite direction (lower fields, or higher frequencies). Such power-dependent shifts of the magnon--polariton branches are the signatures of the MKE. The OOP measurement at \SI{25}{dBm} also reveals a marked foldover effect of the resonance \cite{PhysRev.100.1788, 10.1063/1.334997}, characterized by sharp changes in the transmission of the upper branch near \SI{533}{mT}. Discussions on the foldover effect and bistability are presented in SM \cite{supp}. The ability to observe a strong MKE with positive or negative Kerr shift, selected by the magnetic-field orientation, represents a striking departure from previous measurements~\cite{PhysRevB.94.224410, PhysRevLett.120.057202, PhysRevLett.129.123601, Shen2025}. 

\begin{figure}[t]
\includegraphics[width=\linewidth]{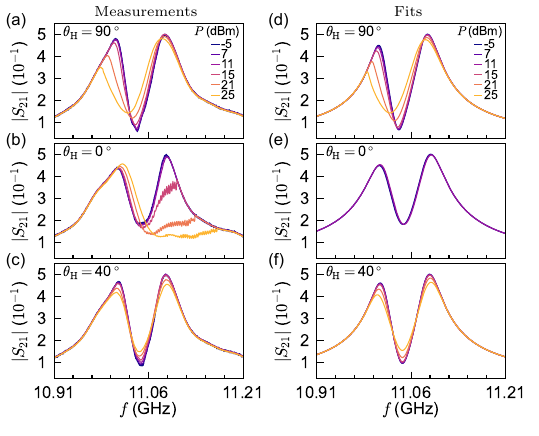}%
\caption{Power- and frequency-dependent microwave transmission for the coupled magnon-photon system at a fixed field near the center of the avoided crossing. Field angle is (a,d) $\theta_\mathrm{H}=\SI{90}{\degree}$, (b,e) $\theta_\mathrm{H}=\SI{0}{\degree}$, and (c,f) $\theta_\mathrm{H}=\SI{40}{\degree}$, for (a,b,c) measurements and (d,e,f) fits.}
\label{fig:fig3}
\end{figure}

For further analysis, we measure $S_{21}$ against frequency at a constant magnetic field near the center of the avoided crossing, and monitor shifts of the two polariton peaks in finer power steps. \Cref{fig:fig3}(a) shows the measurements for IP field ($\theta_\mathrm{H}=\SI{90}{\degree}$). At high powers, the magnon--polariton peaks shift to lower frequencies. The shift is more apparent in the lower branch, which acquires a higher magnon character than the upper branch at high powers \cite{petrosyan2026modelmagnonkerreffect}. The measurement for OOP field ($\theta_\mathrm{H}=\SI{0}{\degree}$) is shown in \cref{fig:fig3}(b), which corresponds to a shift in the opposite direction. Moreover, at \SI{15}{dBm} and above, oscillations with a periodicity of \SI{3.2}{MHz} appear, corresponding to standing bulk acoustic modes in the YIG/GGG stack \cite{PhysRevB.106.014407}. These oscillations appear near the upper magnon--polariton, due to the magnon-like character of this branch.

The sign reversal between the IP and OOP Kerr shifts suggests that at an intermediate angle, the MKE should be suppressed. We identify the angle where the MKE cancels by performing measurements similar to that in \cref{fig:fig3}(a), with \SI{5}{\degree} steps, from IP to OOP configurations. \Cref{fig:fig3}(c) shows the measurement at $\theta_\mathrm{H}=\SI{40}{\degree}$, where the magnon--polariton peaks do not shift in frequency with power. However, the amplitude of the transmission decreases with increasing power due to enhanced damping in the YIG film originating from nonlinear magnon scattering \cite{supp}.

To describe and quantitatively determine the frequency shift due to the MKE in a thin film magnet, we consider the following Hamiltonian for the coupled magnon--photon system \cite{PhysRevB.94.224410, petrosyan2026modelmagnonkerreffect}:
\begin{equation}\small\label{eq:Hamiltonian_full}
    \hat{H}/\hbar = \omega_\mathrm{c} \hat{c}^\dagger \hat{c}+ \omega_\mathrm{m} \hat{m}^\dagger \hat{m} + g(\hat{c}^\dagger \hat{m}+ \hat{c} \hat{m}^\dagger) +  \mathcal{K}\hat{m}^\dagger \hat{m} \hat{m}^\dagger \hat{m},
\end{equation}
where $\omega_\mathrm{c}$ and $\omega_\mathrm{m}$ are angular frequencies of the cavity and Kittel mode, respectively, $\hat{c}^\dagger (\hat{c})$ are the creation (annihilation) operators of the cavity mode, $\hat{m}^\dagger (\hat{m})$ are the creation (annihilation) operators of the Kittel mode, $\mathcal{K}$ is the Kerr coefficient, and $\hbar$ is the reduced Planck's constant. Higher order squeezing terms in the Hamiltonian are neglected under the rotating wave approximation \cite{PhysRevB.94.224410, petrosyan2026modelmagnonkerreffect}. We use the mean field approximation for the quadratic term, similar to previous studies \cite{PhysRevB.94.224410}, and obtain
\begin{equation}\small\label{eq:Hamiltonian_full_MF}
    \hat{H}/\hbar = \omega_\mathrm{c} \hat{c}^\dagger \hat{c}+ (\omega_\mathrm{m} + 2\mathcal{K}\langle \hat{m}^\dagger \hat{m} \rangle) \hat{m}^\dagger \hat{m} + g(\hat{c}^\dagger \hat{m}+ \hat{c} \hat{m}^\dagger).
\end{equation}
Here, $\langle \hat{m}^\dagger \hat{m} \rangle$ is the expectation value for the magnon population and the term $2\mathcal{K}\langle \hat{m}^\dagger \hat{m} \rangle$ modulates the magnon angular frequency, which induces the magnon--polariton shift. 

The equations of motion of this Hamiltonian combined with input-output formalism provide a model for $S_\mathrm{21}$, which we use to fit the experimental data \cite{supp}. \Cref{fig:fig3}(d) and (e) show fits to the data at $\theta_\mathrm{H}=\SI{90}{\degree}$ and $\theta_\mathrm{H}=\SI{0}{\degree}$, respectively, from which we obtain $\mathcal{K}_\mathrm{IP}/(2\pi)=-58\pm\SI{6}{nHz}$ and $\mathcal{K}_\mathrm{OOP}/(2\pi)=170\pm\SI{27}{nHz}$. At $P=\SI{25}{dBm}$, the number of magnons in the system is of the order of \SI{e14}{}. We restrain the fit for OOP field to below \SI{11}{dBm} to avoid the foldover regime and magnon--phonon coupling present at higher powers, which cannot be analyzed by the present input-output model. Finally, \cref{fig:fig3}(f) shows the fit at $\theta_\mathrm{H}=\SI{40}{\degree}$, from which we obtain $\mathcal{K}/(2\pi)=-3.1\pm\SI{0.9}{nHz}$, highlighting the near cancellation of the MKE. 

We now compare the results to theoretical prediction. The Kerr coefficients for a thin film magnetized IP and OOP are given by \cite{petrosyan2026modelmagnonkerreffect}
\begin{equation}\label{eq:Kerr_coeff_IPOOP}
    \mathcal{K}_{\rm{IP}} = \xi \frac{\hbar\gamma^2K_\mathrm{u}^*}{2M_\mathrm{s}^2V_\mathrm{m}}, \quad
    \mathcal{K}_{\rm{OOP}} = -\frac{\hbar\gamma^2K_\mathrm{u}^*}{M_\mathrm{s}^2V_\mathrm{m}},
\end{equation}
where $\gamma$ is the gyromagnetic ratio, $K_\mathrm{u}^*$ the effective anisotropy energy density (including contributions from uniaxial anisotropy mechanisms along the growth direction and shape anisotropy), $M_\mathrm{s}$ the saturation magnetization, $V_\mathrm{m}$ the volume of the film, $\xi=(B_\mathrm{eff}/4+\mu_0 H_\mathrm{ext})/(B_\mathrm{eff}+\mu_0 H_\mathrm{ext})=0.7$ a geometric factor arising from the ellipticity of the magnetic precession \cite{petrosyan2026modelmagnonkerreffect}, and $B_\mathrm{eff}=-2K_\mathrm{u}^*/M_\mathrm{s}$. We neglect the cubic magnetocrystalline anisotropy constant $K_1$ in the Kerr coefficient as it is small compared to $K_\mathrm{u}^*$ in thin-film YIG. The opposite sign of $\mathcal{K}$ in the IP and OOP cases arises from an opposite nonlinearity in the action of the anisotropy field ($\propto -\cos{\theta_{\rm{M}}}$) on the dynamical magnetization (bringing it back to the plane for IP field, or tilting it further away from normal for OOP field), resulting in different directions of Kerr shift as seen in \cref{fig:fig2} and \cref{fig:fig3}. Moreover, $|\mathcal{K}_{\rm{IP}}|$ is at most half of $\mathcal{K}_{\rm{OOP}}$, as the anisotropy field acts during the full precession cycle in the OOP geometry, but only half the cycle in the IP geometry, when the dynamic magnetization has an OOP component.

To estimate $K_\mathrm{u}^*$, we fit the magnon resonance field $\mu_0 H_\mathrm{res}$ versus $\theta_\mathrm{H}$ at $f=\SI{11.055}{GHz}$ and obtain $K_\mathrm{u}^*=\SI{-11.8}{kJ.m^{-3}}$ \cite{supp}, much larger in magnitude than $K_1=\SI{-610}{J.m^{-3}}$ in YIG \cite{Hansen1974a}. Using \cref{eq:Kerr_coeff_IPOOP} with $M_\mathrm{s}=\SI{144}{kA.m^{-1}}$ and $\gamma/(2\pi) = \SI{28}{GHz.T^{-1}}$, we obtain $\mathcal{K}_{\rm{IP}}/(2\pi)=\SI{-60.3}{nHz}$ and $\mathcal{K}_{\rm{OOP}}/(2\pi)=\SI{168.5}{nHz}$, which agrees very well with the experimental values. Notably, these magnon Kerr coefficients are two orders of magnitude larger than those of the smallest YIG spheres, as a result of the smaller magnetic volume and inherently large shape anisotropy in thin films, making them more suitable for inducing nonlinear magnonic phenomena.

\begin{figure}[t]
\includegraphics[width=\linewidth]{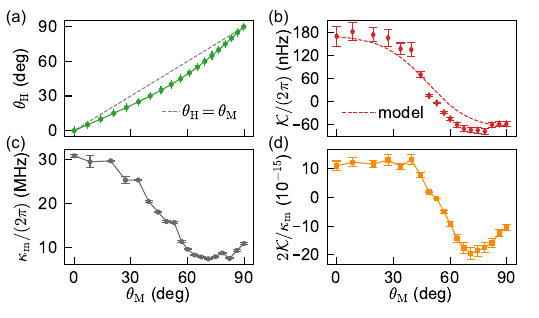}%
\caption{Dependence of (a) field angle $\theta_\mathrm{H}$, (b) Kerr coefficient $\mathcal{K}/(2\pi)$, (c) magnon dissipation rate $\kappa_\mathrm{m}/(2\pi)$, and (d) anharmonicity $2\mathcal{K}/\kappa_\mathrm{m}$, all as functions of $\theta_\mathrm{M}$.}
\label{fig:fig4}
\end{figure}

In \cref{fig:fig4}, we report the complete angle dependence for our measurements. The noncollinear external and anisotropy fields at intermediate $\theta_\mathrm{H}$ between IP and OOP result in a difference between $\theta_\mathrm{H}$ and $\theta_\mathrm{M}$, with the latter being closer to \SI{90}{\degree}. We obtain $\theta_\mathrm{H}$ against $\theta_\mathrm{M}$ from the angle dependence of the Kittel mode \cite{supp} [\cref{fig:fig4}(a)]. \Cref{fig:fig4}(b) shows $\mathcal{K}$ as a function of $\theta_\mathrm{M}$ and highlights the continuous tunability of $\mathcal{K}$ with magnetization angle. 
We model the angle dependence using
\begin{equation}
    \label{eq:theta_Kerr}
    \mathcal{K}=\mathcal{K}_{\rm{OOP}}(\xi''\cos^2\theta_{\rm{M}}-\xi'\sin^2\theta_{\rm{M}}/2),
    \end{equation}
where $\xi'<1$ and $\xi''>1$ are analytical geometric factors extending $\xi$ into the angle-dependent $\mathcal{K}$ \cite{petrosyan2026modelmagnonkerreffect}. Using only $\mathcal{K}_\mathrm{OOP}/(2\pi)=\SI{168.5}{nHz}$ and $\theta_\mathrm{M}$ obtained from \cref{fig:fig4}(a) as input parameters, our model matches well the experimental data [dashed line in \cref{fig:fig4}(b)]. It also predicts the cancellation of the MKE near the magic angle $\theta_{\rm{M}}=\SI{54.7}{\degree}$, in good agreement with the observed cancellation at $\theta_\mathrm{M}=\SI{51.4}{\degree}$ ($\theta_\mathrm{H}=\SI{40}{\degree}$). The remaining discrepancy is attributed to the omission of $K_1$ in the model.

Since the visibility of the Kerr shift depends on the magnitude of $2\mathcal{K}$ relative to the magnon linewidth $\kappa_\mathrm{m}$, we use $2\mathcal{K}/\kappa_\mathrm{m}$ as a figure of merit for the anharmonicity per magnon of the system. The dependence of $\kappa_\mathrm{m}/(2\pi)$ on $\theta_\mathrm{M}$ is shown in \cref{fig:fig4}(c). 
This complex behavior results from the combination of two-magnon scattering and inhomogeneous broadening effects \cite{Soumah2018} and extends further into the angle dependence of $2\mathcal{K}/\kappa_\mathrm{m}$, shown in \cref{fig:fig4}(d). Remarkably, although the Kerr coefficient is largest for OOP magnetization, the largest $2|\mathcal{K}|/\kappa_\mathrm{m}\sim \SI{2e-14}{}$ is obtained neither for IP nor OOP magnetizations but at an intermediate angle near $\theta_\mathrm{M}=\SI{70}{\degree}$ (full transmission data shown in SM~\cite{supp}).
\begin{figure}[t]
\includegraphics[width=\linewidth]{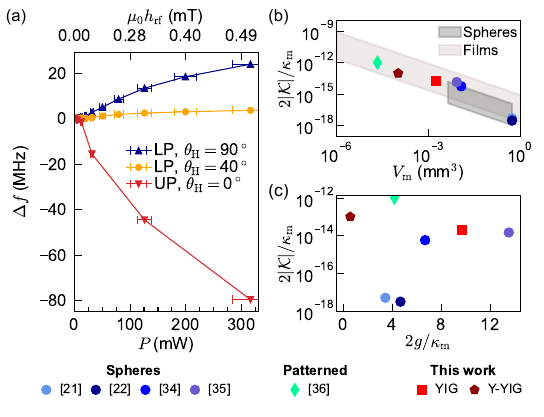}%
\caption{(a) Microwave power- and field-dependence of lower and upper magnon--polariton (LP and UP) shifts obtained from the spectra in \cref{fig:fig3}(a-c). For the UP at $\theta_\mathrm{H}=\SI{0}{\degree}$, we used the frequency before the foldover as the polariton position for $P\geq$~\SI{15}{dBm}. Experimental observations based on previous works \cite{PhysRevB.94.224410, PhysRevLett.120.057202, PhysRevLett.129.123601, Shen2025, zw18-26nw} for (b) anharmonicity $2|\mathcal{K}|/\kappa_\mathrm{m}$ versus magnetic volume for spheres and films and (c) $2|\mathcal{K}|/\kappa_\mathrm{m}$ versus $2g/\kappa_\mathrm{m}$. The shaded region for spheres in (b) represents attainable values of $2|\mathcal{K}|/\kappa_\mathrm{m}$ calculated for YIG, with $\kappa_\mathrm{m}/(2\pi)$ ranging \SI{1}{}--\SI{50}{MHz}. For films, the shaded region is both for IP and OOP fields, calculated for YIG and Y-YIG at $f=\SI{10}{GHz}$, with $\kappa_\mathrm{m}/(2\pi)$ ranging \SI{1}{}--\SI{100}{MHz}.}
\label{fig:fig5}
\end{figure}

To illustrate the consequences of the large $2|\mathcal{K}|/\kappa_\mathrm{m}$, \cref{fig:fig5}(a) shows the shift of the upper and lower magnon--polaritons (UP and LP) obtained from the measured spectra in \cref{fig:fig3}, versus applied microwave power and microwave field \cite{1mc9-k683}. Notably, the LP shift at $\theta_\mathrm{H}=\SI{90}{\degree}$ and the UP shift at $\theta_\mathrm{H}=\SI{0}{\degree}$ exceed their respective magnon dissipation rates. The shift of the UP reaches \SI{80}{MHz}, which places it far away from the initial anticrossing regime, strongly detuned from the bare cavity resonance, and making the cavity transmission conditional on the magnetization dynamics in the YIG film, as was introduced in \cref{fig:fig1_MKE}(c).

An important feature is the universality of the MKE deriving from shape anisotropy, which does not depend on the magnetic material hosting the magnons or on the temperature of the magnetic system. Indeed, when shape anisotropy dominates over magnetocrystalline anisotropy, $K_{\rm{u}}^*=-\mu_0M_{\rm{s}}^2/2$ and \cref{eq:Kerr_coeff_IPOOP} for OOP field yields $\mathcal{K}=\hbar \gamma^2\mu_0/(2V_\mathrm{m})$, a universal parameter that depends only on the sample volume $V_\mathrm{m}$. For IP field, $\mathcal{K}$ is weakly dependent on $M_{\rm{s}}$ (and hence also on temperature), through the ellipticity parameter $\xi$.

Beyond shape anisotropy, a key benefit of thin films is that interfacial effects \cite{PhysRevLett.124.257202, PhysRevLett.130.126703}, stoichiometry \cite{PhysRevMaterials.7.094405, PhysRevMaterials.6.114402, https://doi.org/10.1002/adfm.202503644}, and epitaxial strain \cite{KUBOTA201363,PhysRevB.89.134404, Soumah2018, https://doi.org/10.1002/admi.202300217} can be further adjusted to control the MKE. To demonstrate this, we measure a \SI{35}{nm}-thick, $\SI{2.4}{}\times\SI{1.2}{mm^2}$ yttrium-rich YIG (Y-YIG) film grown epitaxially on GGG \cite{https://doi.org/10.1002/adfm.202503644}. The larger lattice parameter in Y-YIG compared to GGG induces compressive IP strain in the film, enhancing the effective anisotropy energy density $K_\mathrm{u}^*$ by nearly \SI{50}{\%} compared to YIG grown on GGG, see \cite{supp} for details. By performing measurements similar to \cref{fig:fig3} in the IP configuration, we now obtain $\mathcal{K}_\mathrm{IP}/(2\pi)=-1.6\pm\SI{0.3}{\micro Hz}$, and anharmonicity reaching $10^{-13}$ [\cref{fig:fig5}(b)]. The significant improvement in $\mathcal{K}_\mathrm{IP}$ arises from the combined effects of reduced volume in this smaller and thinner sample, and stronger easy-plane anisotropy.

We compare in \cref{fig:fig5}(b) the experimental results for the MKE in our thin films and YIG spheres from previous studies. Reported values of Kerr coefficient for spheres range in the orders of $\SI{0.01}{nHz}$--$\SI{1}{nHz}$, largely determined by their volume \cite{PhysRevB.94.224410,PhysRevLett.120.057202, PhysRevLett.129.123601, Shen2025}. In our extended Y-YIG thin film at room temperature, $|\mathcal{K}|/(2\pi)=\SI{1.6}{\micro Hz}$ is already three orders of magnitude higher than that of the smallest spheres, and $2|\mathcal{K}|/\kappa_{m}$ is an order of magnitude larger despite the stronger magnon dissipation rate occurring in this sputtered film. The detail of the magnetic parameters entering this analysis and further comparison against spheres or patterned thin-film disks with ideal magnetic properties are provided in SM~\cite{supp}.
 
In perspective, patterned thin-film magnets allow for a much larger reduction in volume compared to bulk systems, easily reaching values as small as $\sim$ \SI{e-9}{mm^3}. This reduction favors much larger $2|\mathcal{K}|/\kappa_\mathrm{m}$ ratios, while also enabling sign tunability between IP and OOP fields, further control of magnetic anisotropy as demonstrated above, and extension to antiferromagnets \cite{Boventer2023,Wang2025c}. Patterning would also mitigate the increase in $\kappa_\mathrm{m}$ due to inhomogeneous broadening. Reaching the limit of $V_\mathrm{m}\sim\SI{e-16}{mm^3}$ in YIG, with $\sim5000$ spins, would result in $2|\mathcal{K}|\sim\kappa_\mathrm{m}$, implementing an effective two-level system as required for obtaining a qubit behavior \cite{PhysRevB.102.100402}.

The trade-off for enhancing the anharmonicity of the magnonic system by reducing its volume is the coherence of the magnon--photon interaction, i.e., the magnon--photon coupling constant $g$ relative to the magnon half-width-at-half-maximum linewidth $\kappa_\mathrm{m}/2$. Since $g\sim{}g_{\rm{s}}\sqrt{N}$, with $g_{\rm{s}}$ the single-spin coupling from the resonator, the condition $g>\kappa_\mathrm{m}/2$ requires a sufficiently large magnonic system and a low magnon linewidth. \Cref{fig:fig5}(c) shows experimental values of $2|\mathcal{K}|/\kappa_\mathrm{m}$ versus $2g/\kappa_\mathrm{m}$ for YIG spheres and thin films. Our extended YIG film coupled to the loop-gap resonator provides large $2g/\kappa_\mathrm{m}$ despite a lower number of spins compared to the spheres. Integration of a small volume of patterned magnetic film with a lumped-element resonator therefore appears as a most suitable route to realize strongly anharmonic hybrid systems, potentially leading to the magnon-antibunching regime.

In summary, our investigation of the MKE in magnetic insulator thin films coupled to a microwave resonator demonstrates large magnon Kerr shifts, that can be continuously tuned in magnitude and direction by controlling the equilibrium magnetization angle. We further show the enhancement of the MKE with reduced volume and stronger effective anisotropy energy of a strained film. Our experimental data and theoretical modeling provide a quantitative description of the Kerr non-linearity in magnetic thin films, rationalizing the enhancement or suppression of the MKE depending on sample and experimental parameters. The MKE in extended films is orders of magnitude larger than in YIG spheres. Extending the present results to patterned thin-film magnets offers the prospect of massive and fully tunable anharmonicity, promising compact non-linear cavity-magnonic devices.

\section*{Acknowledgments} This work was supported by the Swiss National Science Foundation (Grant No.~10004526). We acknowledge CNRS Physics for supporting our work through an international research project (IRP 2025 CORMORANT). H.M.~acknowledges support from JSPS Postdoctoral Fellowship (Grant No.~23KJ1159). H.W.~acknowledges the support of the China Scholarship Council (CSC, Grant No.~202206020091). R.S.~acknowledges funding by the Deutsche Forschungsgemeinschaft via the SFB 1432, Project No.~425217212. W.L.~acknowledges the support of the ETH Zurich Postdoctoral Fellowship Program (21-1 FEL-48) and from a government grant managed by the Agence Nationale de la Recherche as part of the France 2030 program, with reference ANR-24-EXSP-0005 (“MAGNON-BRAQET”). 

\section*{Data availability}
The data that support the findings of this article are openly available from the ETH Zurich Research Collection \cite{Petrosyan2026data_MKE_exp}.

%\bibliography{references}
%\clearpage

%\bibliography{ref}

\clearpage
\foreach \p in {1,...,13}{%
  \clearpage
  \includepdf[pages=\p]{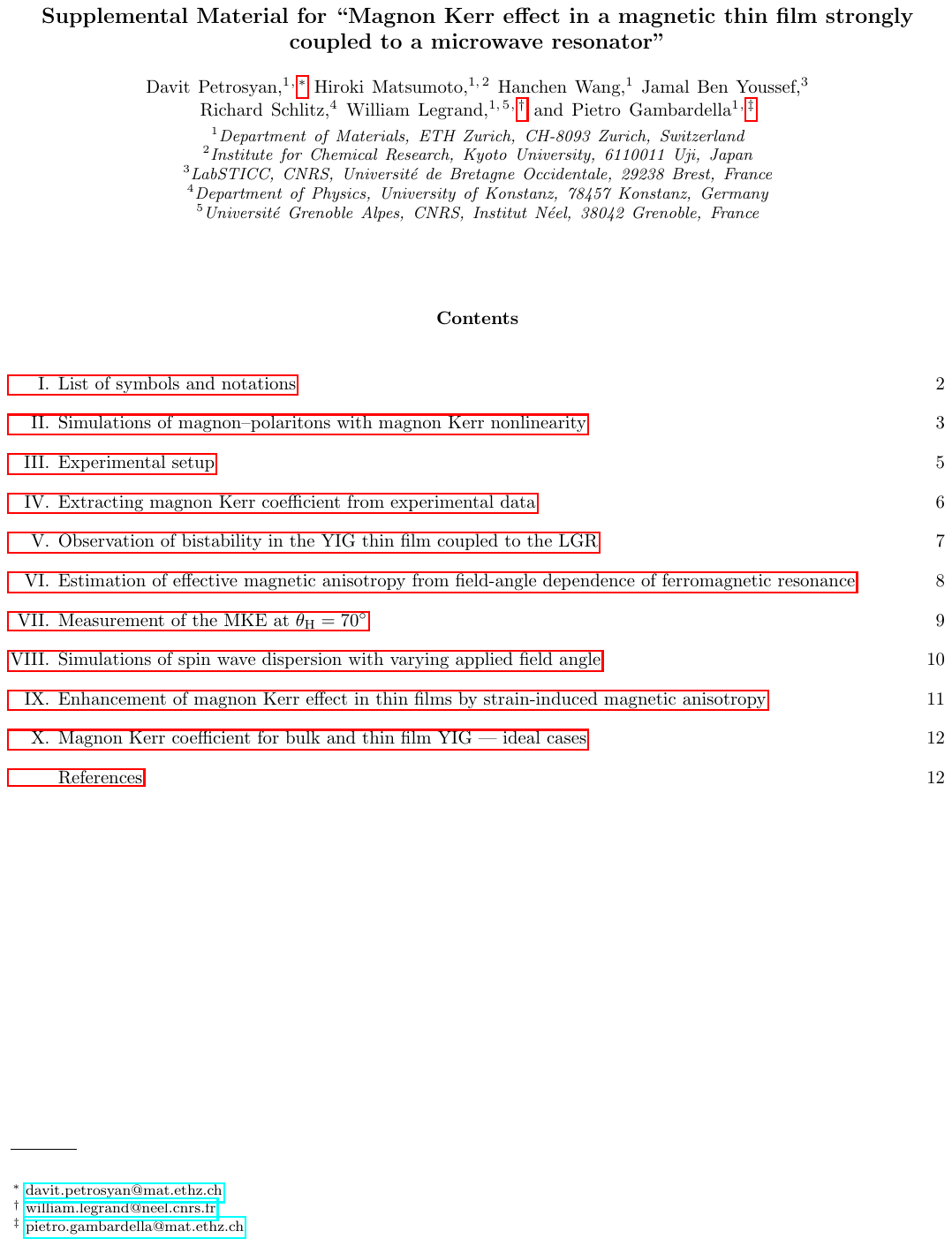}%
}

\end{document}